\documentclass[aip,apl,reprint,a4paper]{revtex4-1}
\usepackage[ansinew]{inputenc}
\usepackage{amsmath}
\usepackage{amsfonts}
\usepackage{amssymb}
\usepackage{graphicx}
\usepackage{siunitx}
\usepackage{comment}

\begin{document}
\title{Nanoscale constrictions in superconducting coplanar waveguide resonators}
\date{\today}
\author{Mark David Jenkins}
\affiliation{Instituto de Ciencia de Materiales de Arag\'on, CSIC - Universidad de Zaragoza, 50009 Zaragoza, Spain}
\affiliation{Departamento de F\'{\i}sica de la Materia Condensada, Universidad de Zaragoza, 50009 Zaragoza, Spain}
\author{Uta  Naether}
\affiliation{Instituto de Ciencia de Materiales de Arag\'on, CSIC - Universidad de Zaragoza, 50009 Zaragoza, Spain}
\affiliation{Departamento de F\'{\i}sica de la Materia Condensada, Universidad de Zaragoza, 50009 Zaragoza, Spain}
\author{Miguel Ciria}
\affiliation{Instituto de Ciencia de Materiales de Arag\'on, CSIC - Universidad de Zaragoza, 50009 Zaragoza, Spain}
\affiliation{Departamento de F\'{\i}sica de la Materia Condensada, Universidad de Zaragoza, 50009 Zaragoza, Spain}
\author{Javier Ses\'e}
\affiliation{Instituto de Nanociencia de Arag\'on, Universidad de Zaragoza E-50009 Zaragoza, Spain}
\affiliation{Departamento de F\'{\i}sica de la Materia Condensada, Universidad de Zaragoza, 50009 Zaragoza, Spain}
\author{James Atkinson}
\affiliation{Department of Physics, University of Central Florida, Orlando, FL 32816, USA}
\author{Carlos S\'anchez-Azqueta}
\affiliation{Dpto. de Ingenier\'{\i}a Electr\'onica y Telecomunicaciones, Universidad de Zaragoza, 50009 Zaragoza, Spain}
\author{Enrique del Barco}
\affiliation{Department of Physics, University of Central Florida, Orlando, FL 32816, USA}
\author{Johannes Majer}
\affiliation{Vienna Center for Quantum Science and Technology, Atominstitut, TU Wien, 1020 Vienna, Austria}
\author{David Zueco}
\affiliation{Instituto de Ciencia de Materiales de Arag\'on, CSIC - Universidad de Zaragoza, 50009 Zaragoza, Spain}
\affiliation{Departamento de F\'{\i}sica de la Materia Condensada, Universidad de Zaragoza, 50009 Zaragoza, Spain}
\author{Fernando Luis}
\email{fluis@unizar.es} \affiliation{Instituto de Ciencia de Materiales de Arag\'on, CSIC - Universidad de Zaragoza, 50009 Zaragoza, Spain}
\affiliation{Departamento de F\'{\i}sica de la Materia Condensada, Universidad de Zaragoza, 50009 Zaragoza, Spain}

\begin{abstract}
We report on the design, fabrication and characterization of superconducting coplanar waveguide resonators with nanoscopic constrictions. By reducing the size of the center line down to $50$ nm, the radio frequency currents are concentrated and the magnetic field in its vicinity is increased. The device characteristics are only slightly modified by the constrictions, with changes in resonance frequency lower than $1$ \% and internal quality factors of the same order of magnitude as the original ones. These devices could enable the achievement of higher couplings to small magnetic samples or even to single molecular spins and have applications in circuit quantum electrodynamics, quantum computing and electron paramagnetic resonance.
\end{abstract}
\maketitle

The field of cavity quantum electrodynamics (QED) studies the interaction of photons in resonant cavities with either natural or "artificial" atoms, such as quantum dots and superconducting qubits, having a nonlinear and discrete energy level spectrum.\cite{Mabuchi2002,Schoelkopf2008} For applications in spectroscopy and especially quantum information processing a major goal is to maximize the coupling strength $g$ of the atom to either electric or magnetic cavity fields, making it larger than the decoherence rates of both the cavity and the atom (strong coupling regime).

\begin{figure}
\includegraphics[width=\columnwidth]{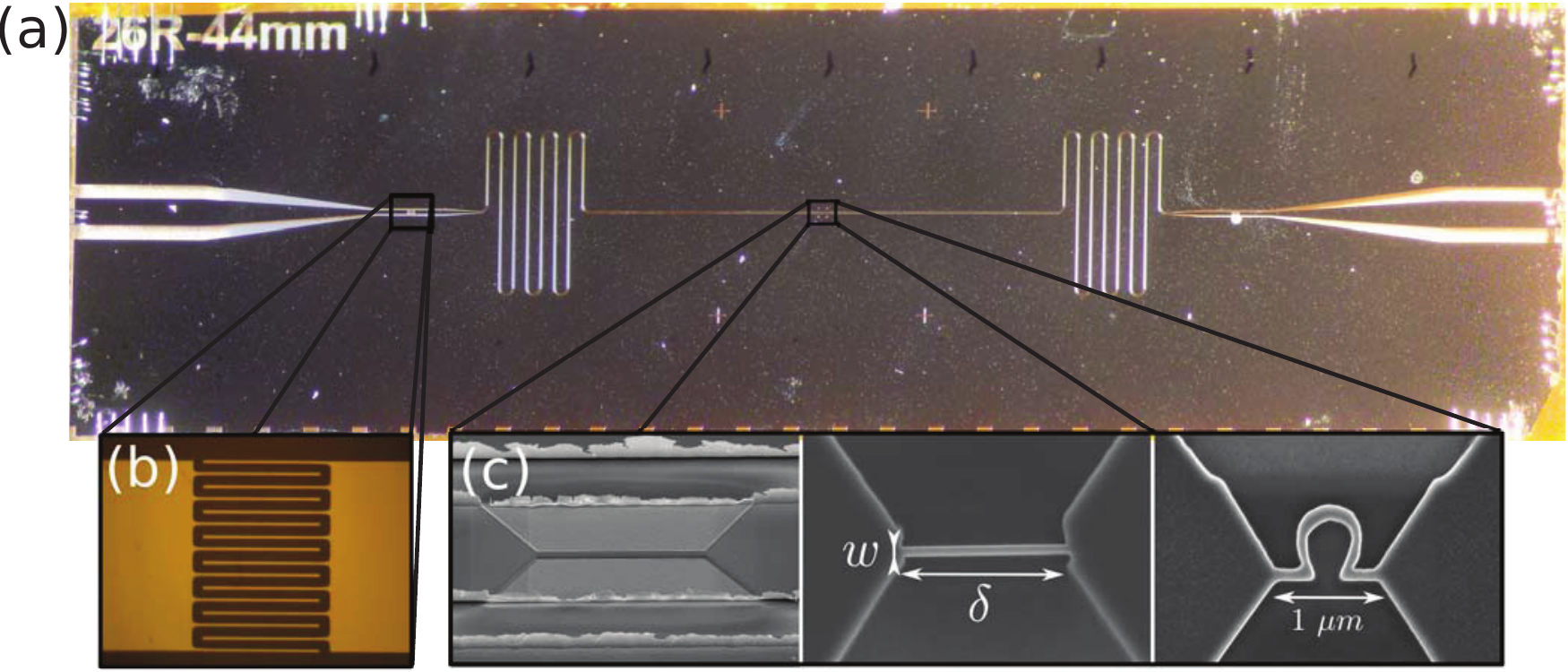}
\caption{(a) Microscope photograph of a coplanar waveguide resonator fabricated with niobium deposited on a sapphire substrate. The length of the resonator segment is \SI{44}{\milli\metre}, which corresponds to a resonance frequency $f_{0} \simeq 1.3-\SI{1.4}{\giga\hertz}$. (b) Close-up of a gap coupling capacitor. (c) Scanning electron microscope images of three nano-constrictions fabricated in the center line of the resonator.}
\label{fig:CPWG}
\end{figure}

One of the most direct ways of enhancing $g$ is to reduce the cavity size and therefore increase the electromagnetic energy density and field strength at the atom site. This is the idea behind circuit QED,\cite{Blais2004} in which the classical three dimensional resonant cavities are replaced by sections of microwave transmission lines. The prime example of this type of cavity is a coplanar waveguide (CPW) resonator, i.e., a section of coplanar waveguide capacitively coupled to external feed lines (see Fig. \ref{fig:CPWG}). As a first approximation, these systems can be considered as one dimensional resonators, where the resonant frequency $f_{0}$ is controlled by the length of the transmission line segment. The photon energy is concentrated in and around the center line, leading to field strengths up to $100$ times larger than in typical three dimensional cavities.\cite{Blais2004} Furthermore, if the lines are superconducting the resistive losses can be suppressed to achieve quality factors $Q$ of up to $10^6$. In these cases, the losses and dephasing of the composite system are often limited only by the atom properties.

Previous studies have shown the performance of these devices,\cite{Frunzio2005,Goppl2008} and their strong coupling with different types of quantum two-level systems, such as superconducting qubits\cite{Wallraff2004,Majer2007,Abdumalikov2008,Dicarlo2010,Niemczyk2010} or quantum dots,\cite{Petersson2012} at the single photon level. Strong coupling has also been achieved to collective spin states of, e.g., nitrogen vacancy centers in diamond and other magnetic systems.\cite{Schuster2010,Kubo2010a,Chiorescu2010,Amsuss2011} Electron spins are attractive due to their usually longer coherence times, as compared with electrical degrees of freedom, which could allow longer storage times for quantum information.\cite{Schoelkopf2008} Strong coupling to single spins or small ensembles of them has not been achieved yet, due to their weaker coupling to the electromagnetic radiation. Achieving this limit would open new possibilities in quantum information and related fields. Artificial molecular magnets, synthesized by chemical methods, are of especial interest as they provide realizations of well-defined and identical qubits and quantum logic gates. \cite{Troiani2005,Wedge2012,Martinez-Perez2012,Luis2011} In a previous work,\cite{Jenkins2013} it was proposed that strong magnetic coupling could be achieved, even for single molecular magnets, by narrowing the center line of a resonator down to nanometer length scales, as long as the resonator characteristics can be maintained. A similar procedure has been proposed to couple superconducting resonators to the spins of donor defects in silicon \cite{Tosi2014} and to flux qubits.\cite{Abdumalikov2008} In this work we design, fabricate and test superconducting CPW resonators with this type of constrictions and show that their characteristics are stable within a relatively broad range of constriction geometries.

Our devices are fabricated on \SI{500}{\micro\metre} thick C-plane sapphire wafers and consist of a $150$ nm thick niobium layer deposited by radio frequency (RF) sputtering and then patterned by either photolithography and lift-off or reactive ion etching. Nanoscale constrictions were made at the midpoint of the center line by etching it with a focused beam of Ga$^{+}$ ions, using a commercial dual beam system. The ionic current was kept below \SI{20}{\pico\ampere} to maximize the resolution in the fabrication process and to minimize the Nb thickness, of order $10-15$ nm, that is implanted with Ga.\cite{Hao2009,Castan-Guerrero2014} Images of these constrictions, as those shown in Fig. \ref{fig:CPWG}, were obtained in situ by scanning electron microscopy. The microwave transmission measurements were done using a programmable network analyzer at $4.2$ K by mounting the devices on a home-made probe and submerging them in liquid helium.

\begin{figure}
\centering
\includegraphics[width=\columnwidth]{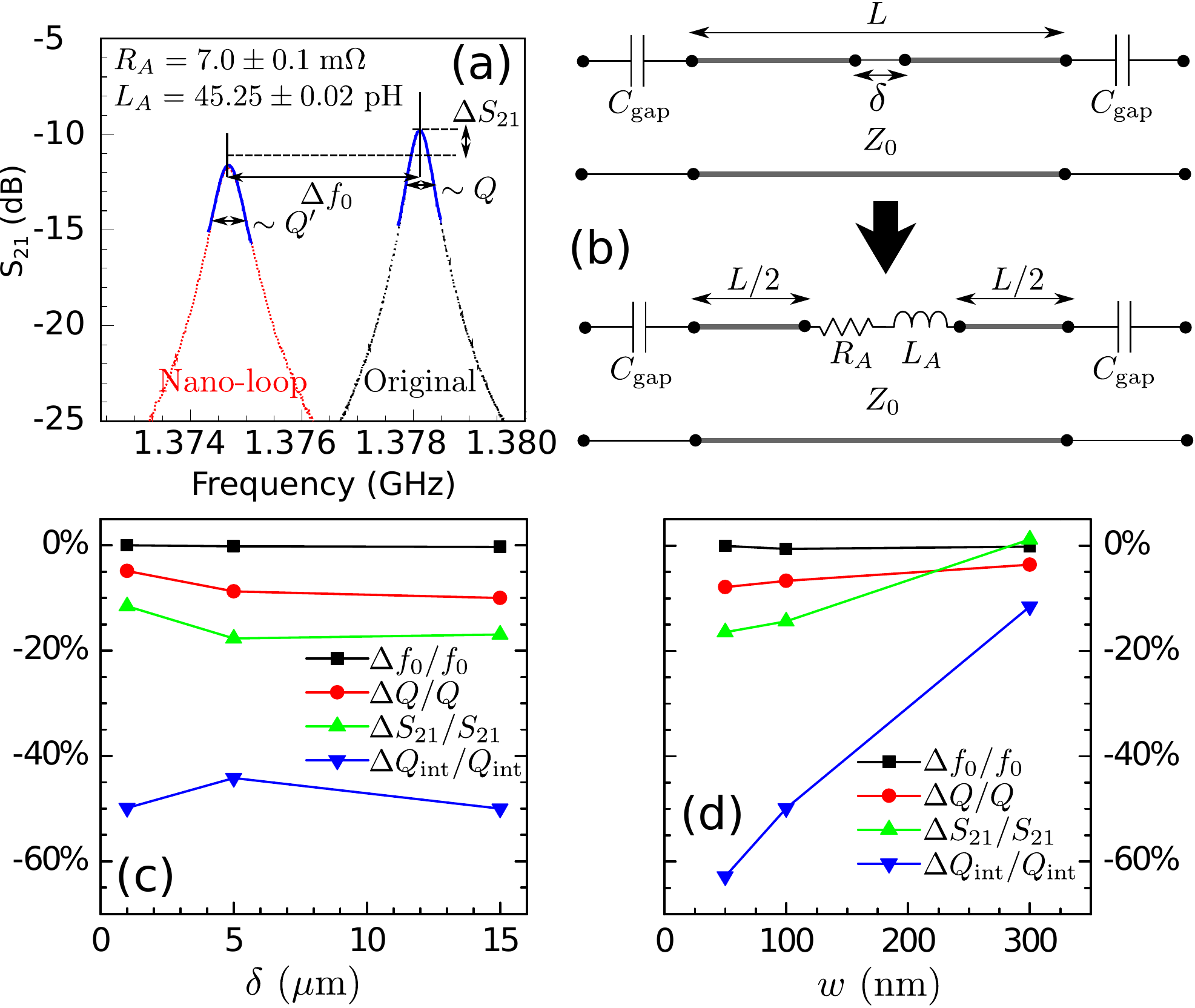}
\caption{(a) Comparison of transmission resonances measured on a superconducting resonator before and after the fabrication of a nanoloop constriction at its center line. $\Delta f_0$, $\Delta Q$ and $\Delta T$ are the variations found in the resonance frequency, quality factor and maximum transmission (in linear scale), respectively. In this particular case, the original parameters were $f_0=\SI{1.378}{\giga\hertz}$ and $Q=2700$ while $\Delta f_0/f_0 = -0.25\%$, $\Delta Q/Q = -22.7\%$ and $\Delta S_{21}/S_{21} = -36\%$. Thick solid lines are least square fits made with the circuit model shown in (b). The presence of a constriction can be taken into account by introducing two effective lump elements $L_{\rm A}$ and $R_{\rm A}$ that account, respectively, for the reduced $f_{0}$ and the enhancement of electromagnetic losses. (c) Variation of resonance parameters for \SI{100}{\nano\metre} wide constrictions of varying lengths (with original values of $f_0\simeq \SI{1.35}{\giga\hertz}$ and $Q\simeq 800$). (d) Variation of resonance parameters for \SI{1}{\micro\metre} long constrictions of varying widths (with original values of $f_0\simeq \SI{1.31}{\giga\hertz}$ and $Q\simeq 250$).}
\label{fig:comp}
\end{figure}

The circuits consist of a large \SI{400}{\micro\metre} center line and \SI{200}{\micro\metre} gaps that narrow down to around \SI{14}{\micro\metre} and \SI{7}{\micro\metre}, respectively after going through gap capacitors. Several types of gap capacitors with a finger design (see Fig. \ref{fig:CPWG}) were fabricated to allow for differently coupled systems. The gaps between the fingers are of \SI{4}{\micro\meter} while the finger lengths are of \SI{100}{\micro\metre}. The length of the cavity is chosen to be $44$ mm by making the waveguide meander across the surface.  Although sapphire has an anisotropic dielectric constant, taking an average value of $\epsilon_r = 10$ is sufficient for our circuit calculations. These parameters give a waveguide characteristic impedance $Z_{0} \simeq$ \SI{50}{\ohm} and an unloaded $f_{0} \simeq \SI{1.5}{\giga\hertz}$. The resonances move to higher frequencies (from \SI{1.3}{\giga\hertz} to \SI{1.405}{\giga\hertz}) and have higher $Q$ (from $100$ to $30000$) the less coupled they are to the feed lines.\cite{suppl} The gap capacitances $C_{\rm gap}$ have been estimated, as illustrated in Fig. \ref{fig:comp}, by fitting the transmission $S_{21}$ {\em vs} frequency data with a simple circuit model, which consists of a segment of transmission line with two identical capacitors connected to the input and output ports. They range from $70$ to \SI{5}{\femto\farad}, of the order of magnitude of values reported in the literature.\cite{Goppl2008}

After the devices were characterized, the center line was narrowed down from about \SI{10}{\micro\metre} to minimum widths $w$ of \SI{50}{\nano\metre} along distances $\delta$ of up to \SI{50}{\micro\metre} (see Fig. \ref{fig:CPWG}). Once the constrictions are made, special care must be taken both with the device manipulation and with the scanning electron microscope imaging since electrostatic buildup and discharge can easily destroy the nanowires. The present study uses two series of three identical resonators. In each series, we make constrictions varying either $\delta$ or $w$ and keeping the other constant. The first series has \SI{100}{\nano\metre} wide constrictions and lengths of \SI{1}{\micro\metre}, \SI{5}{\micro\metre} and \SI{15}{\micro\metre}. Similarly, the second series keeps a constant $\delta = \SI{1}{\micro\metre}$ while $w$ is varied through \SI{300}{\nano\metre}, \SI{100}{\nano\metre} and \SI{50}{\nano\metre}. Other geometries, such as loop constrictions, can also be obtained as we show in Fig. \ref{fig:CPWG}(c). The performances of these devices are analyzed by comparing values of $f_{0}$, $Q$ and maximum $S_{21}$ measured before and after the constrictions were made. The internal quality factor $Q_{\rm int}$, which parameterizes the intrinsic losses of the resonator, has been estimated from the insertion loss, determined from $Q$ and $S_{21}$ as described in [\onlinecite{Goppl2008}]. The values found are shown in Figs. \ref{fig:comp} (c) and (d). Although all the results we present are from transmission measurements ($S_{21},S_{12}$), the same behavior can be seen in the reflection signals ($S_{11},S_{22}$).

The decrease of $f_{0}$, of order $1$\%, likely results from the enhancement of the inductance at the constriction. A simple approach is to model the constriction by a region of length $\delta$ and effective inductance per unit length, $l^{\prime}$, larger than its value $l$ outside this region. As it is described in detail in [\onlinecite{suppl}], this effect makes the center line effectively longer for the propagation of RF currents and leads to a close to linear decrease of $f_{0}$ with increasing $\delta$
\begin{equation}
\frac{\Delta f_{0}}{f_{0}} \simeq -\frac{\delta}{L}\left( \frac{l^{\prime}}{l} - 1 \right)
\label{fvsdelta}
\end{equation}
The experimental data shown in Fig. \ref{fig:comp}(c) are compatible with $l^\prime/l = 10.9$. The decrease of $Q$ and $S_{21}$ indicate that the constriction introduces some extra losses into the system. This is confirmed by the stronger relative variation seen in $Q_{\rm int}$. The constriction constitutes a defect for the propagation of electromagnetic radiation and therefore might enhance reflection to the source. As shown in Fig. \ref{fig:comp}, the experimental results can be accounted for by a circuit model with two additional lumped elements, an inductance $L_{\rm A}$ that accounts for changes of $f_{0}$, and a resistance $R_{\rm A}$ that accounts for $\Delta Q$. From fits such as those shown in Fig. \ref{fig:comp}(a), we find $L_{\rm A}$ to lie in the range of a few tens pH, increasing with $\delta$ as expected from the above considerations, whereas the effective resistance $R_{\rm A}$ is of the order of a few m$\Omega$'s and fairly independent of $\delta$. Despite these additional losses, $Q$ remains mainly limited by the coupling capacitors $C_{\rm g}$ for all devices studied in this work (see Fig. 1 in [\onlinecite{suppl}]). These results show therefore that the fabrication of narrow constrictions does not preclude the attainment of $Q_{\rm int}$ values well above $10^{4}$.

\begin{figure}
\centering
\includegraphics[width=\columnwidth]{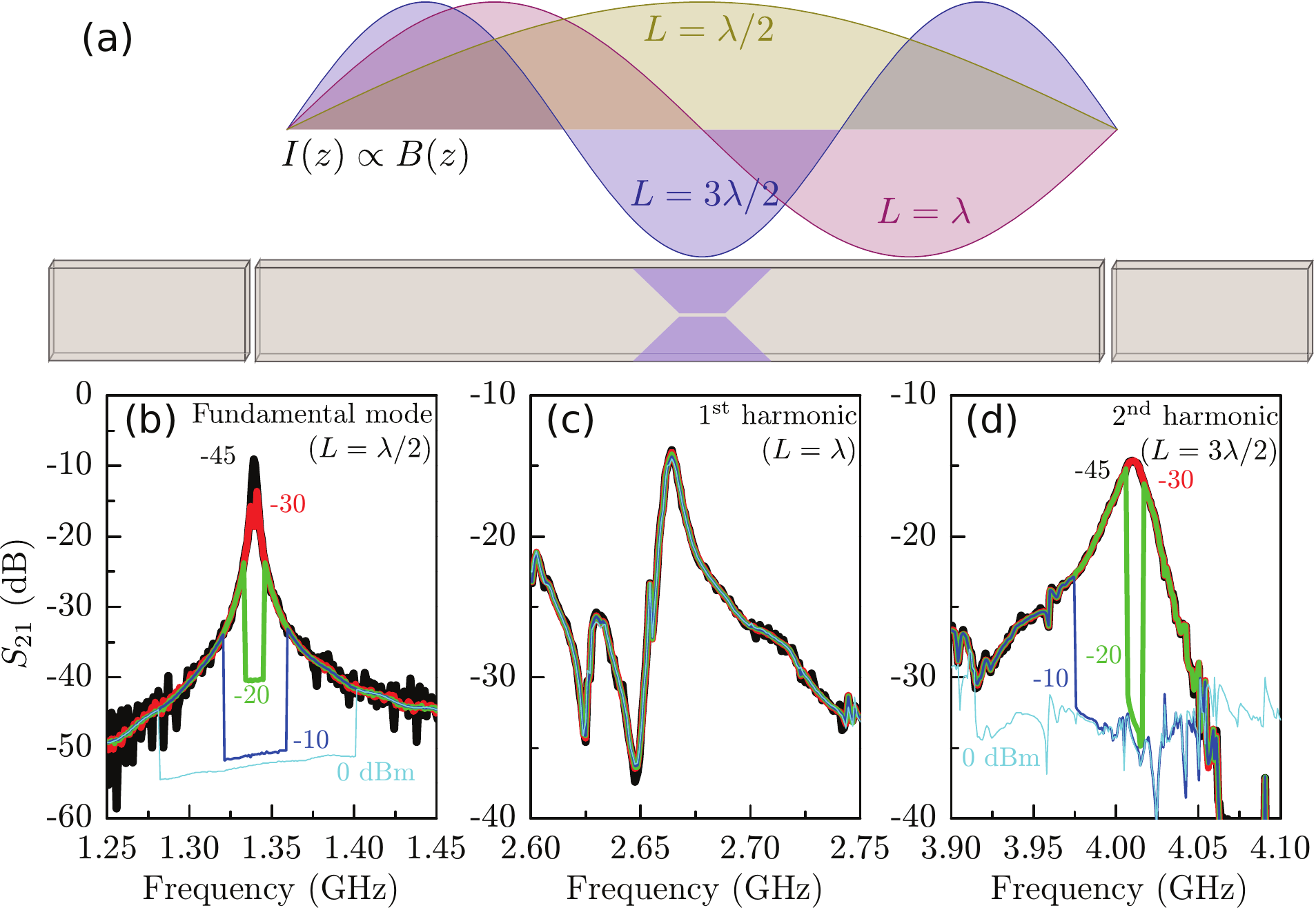}
\caption{Diagram (a) schematically shows the center line of a resonator and the standing current waves for the first three resonant modes. Graphs (b), (c) and (d) show the transmission spectra in a constricted resonator for these three modes and for increasing excitation power. Modes with a current maximum at the constriction (fundamental and second harmonic modes) show a loss of resonance when power is increased while modes with no current at the constriction (first harmonic) show very small changes.}
\label{fig:power}
\end{figure}

Another difference observed in resonators with constrictions is the power dependence of the resonances. Figure \ref{fig:power} shows the transmission through a \SI{100}{\nano\metre} wide and \SI{1}{\micro\metre} long constriction at different excitation powers and for the first three cavity modes. The fundamental mode (Fig. \ref{fig:power}(b)) and the second harmonic (Fig. \ref{fig:power}(d)) break down when power is increased. This effect can be explained by noting that, since both the fundamental mode and second harmonic have a standing wave with a current maximum at the center (as shown by Fig. \ref{fig:power}(a)) where the cross section has been drastically reduced, the constriction can eventually become resistive. By contrast, we find no such effect in the resonance associated with the first harmonic since there is almost zero current flowing through the constriction for this mode. These qualitative arguments can be made quantitative by comparing the current density at which the resonances break down to the critical current of superconducting niobium. The current flowing through the nanowire can be estimated by applying the definition of $Q$ and formulas for the equivalent RLC circuit for the resonator.\cite{Goppl2008} The following expression gives the current amplitude in the resonator
\begin{equation}
I=\sqrt{\frac{\pi Q P_{\rm loss}}{Z_0}},
\end{equation}
\noindent where $P_{\rm loss}$ is the power loss from the resonator (dissipated or lost to the feed lines). From the maximum RF power for which the resonance shown in Fig. \ref{fig:power} remains stable we get a critical current density $j_{\rm c} \simeq \SI{1.4(1)e7}{\ampere\per\centi\metre\squared}$ (corresponding to about \SI{9e9} photons stored in the cavity), which is of the order of magnitude of that found for niobium thin films.\cite{Huebener1975} Experiments performed on resonators with \SI{50}{\nano\metre} wide constrictions show a similar power dependence and give $j_{\rm c}\simeq\SI{1.0(3)e7}{\ampere\per\centi\metre\squared}$.\cite{suppl} These results suggest that the the implantation of Ga at the constriction edges, which is consubstantial to focussed ion beam lithography, does not dramatically affect the superconductivity of Nb. They also evidence that the current is forced to flow through the constrictions and that its density, thus also the RF magnetic field, is being locally enhanced.

\begin{figure}
\centering
\includegraphics[width=\columnwidth]{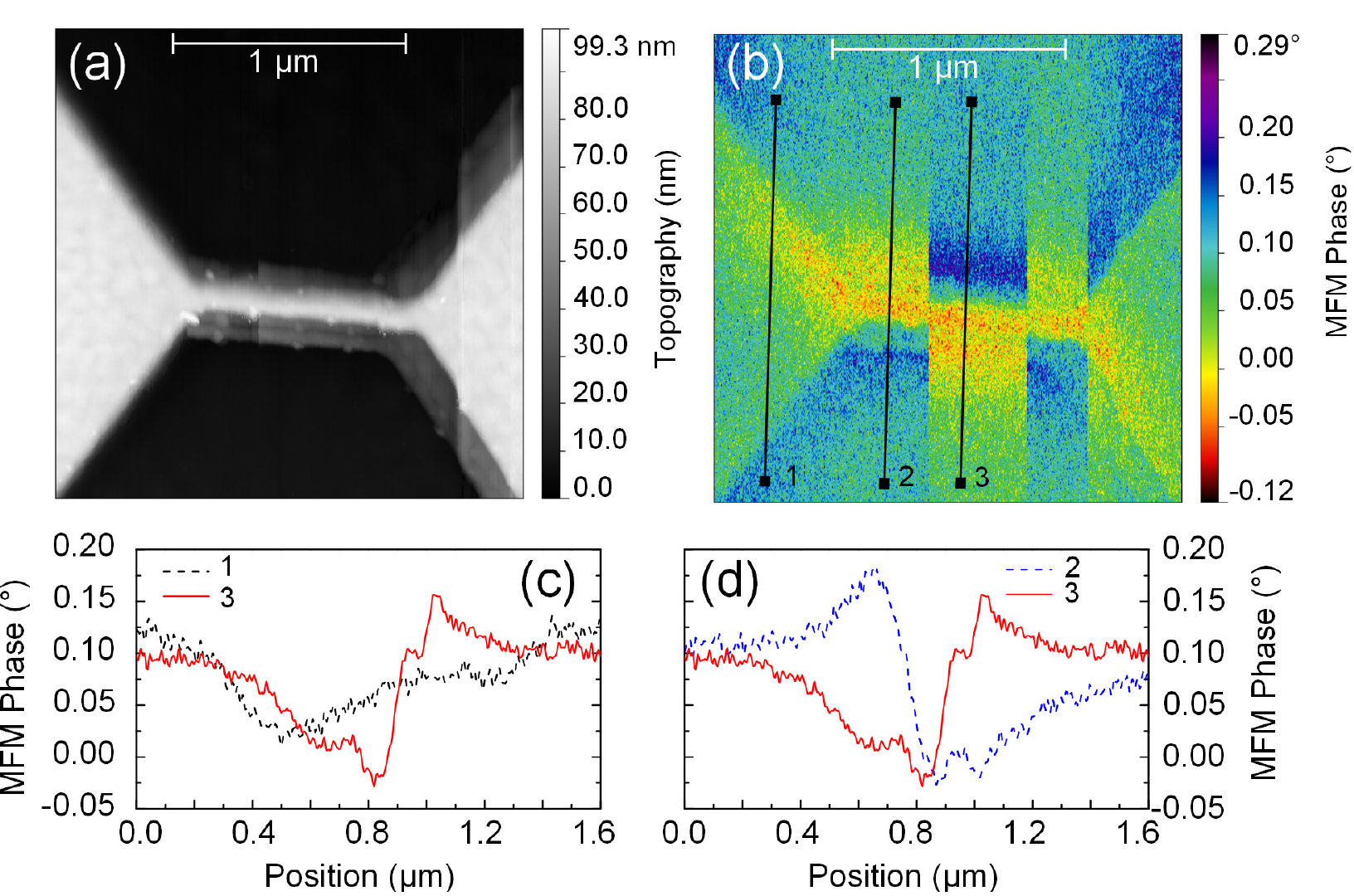}
\caption{(a) Atomic force microscopy image of a \SI{1}{\micro\metre} by \SI{100}{\nano\meter} constriction made from a \SI{100}{\nano\metre} thick gold layer deposited on a sapphire substrate. A \SI{2.1}{\milli\ampere} direct current was flowing through the constriction. (b) MFM signal. The sharp contrast changes in the MFM image correspond to current flips from positive to negative values made during the image acquisition. (c) and (d) MFM profiles taken at three different locations. A clear magnetic signal is measured near the wire, which is not visible in the wider area.}
\label{fig:afm}
\end{figure}

In order to explore such enhancement in a more direct manner, we locally measured, by means of magnetic force microscopy (MFM), the magnetic field generated by a current flowing via one constriction. A \SI{1}{\micro\metre} long by \SI{100}{\nano\meter} wide constriction was fabricated out of a \SI{100}{\nano\metre} thick gold layer deposited on sapphire. The circuit was then mounted on the MFM stage and connected to a direct current source at room temperature. MFM images of the constriction area were measured while a \SI{2.1}{\milli\ampere} current was flowing through it. The resulting topographic and magnetic phase images are shown in Fig. \ref{fig:afm}. These images show a clear magnetic signal (a magnetic phase contrast of $\Delta \phi = 0.26^\circ$ from maximum to minimum) in the constriction area and a negligible signal (indistinguishable from the background) in the wider areas of the circuit. Also, the current was reversed three times during the image acquisition. Each time, a sharp contrast change was observed in the phase image, thus showing that the signal must be magnetic in origin and due to the circulating current. We observe a constant background signal probably due to an electrostatic potential difference between the metallized and non-metallized areas. If greater precision were needed, this background could be filtered out using schemes similar to those detailed in \cite{Yongsunthon2001} where electric potential nulling was used.

We conclude that the fabrication of nanoscopic constrictions in the center line of coplanar superconducting resonators provides a simple and efficient method to locally concentrate the RF magnetic field, with a minor cost in terms of quality factor. Although the resonator frequencies explored in this work are relatively low (\SI{1.3} to \SI{1.4}{\giga\hertz}), there is no reason why this method should not also apply to higher frequencies. Numerical simulations show that the magnetic coupling to individual spins located in the close neighborhood of such constrictions is enhanced by nearly two orders of magnitude, reaching values of order $0.1$ MHz for some single molecule magnets.\cite{Jenkins2013} These devices could potentially be used to develop high sensitivity spectrometers for the characterization of nanoscopic magnetic samples and, provided that sufficiently long spin decoherence times are attained, to achieve strong coupling to individual magnetic qubits.

\begin{acknowledgments}
We acknowledge fruitful discussions with J. J. Garc\'{\i}a-Ripoll and S. Celma. The present work has been partly funded by grants MAT2012-38318-C03, MAT2012-31309, and FIS2011-25167 from the Spanish MINECO, DGA grants E98-MOLCHIP, E19-GEFENOL, and E81, EU project PROMISCE, and Fondo Social Europeo.  Mark D. Jenkins work was funded by a JAE-predoc CSIC grant.
\end{acknowledgments}

\bibliographystyle{apsrev4-1}
\bibliography{paper_CPWG_resub}
\end{document}